\documentclass[num-refs]{wiley-article}

\usepackage{siunitx}
\usepackage{float}
\usepackage{paralist}
\usepackage{tabularx}			
\usepackage{booktabs}			
\makeatletter
\newcommand\notsotiny{\@setfontsize\notsotiny\@vipt\@viipt}
\makeatother

\papertype{Working Paper}
\paperfield{V1.3 --- 27th April 2020}

\title{\Large The \textit{CoRisk}-Index: A data-mining approach to identify industry-specific risk assessments related to COVID-19 in real-time} 

\author[1,2\authfn{1}]{\large Fabian Stephany}
\author[ \authfn{1}]{\large Niklas Stoehr}
\author[3\authfn{1}]{\large Philipp Darius}
\author[4\authfn{1}]{\hspace{3em}\large Leonie Neuh\"auser}
\author[1,4\authfn{1}]{\large Ole Teutloff}
\author[1,5\authfn{1}]{\large Fabian Braesemann}

\contrib[\authfn{1}]{Equally contributing authors. More information about the authors: \url{http://oxford.berlin}}

\affil[1]{Oxford Internet Institute,\hspace{6em}University of Oxford}
\affil[2]{Humboldt Institute for Internet and Society Berlin}
\affil[3]{Centre for Digital Governance,\hspace{6em}Hertie School Berlin}
\affil[4]{Data Science Lab, Hertie School Berlin}
\affil[5]{Sa\"id Business School, University of Oxford}

\corraddress{Fabian Braesemann\\[0.5ex]Sa\"id Business School, University of Oxford, OX1 1HP Oxford, United Kingdom\vspace{0.5ex}}
\corremail{fabian.braesemann@sbs.ox.ac.uk}

\fundinginfo{PD and LN are funded by a PhD scholarship of the Hertie Foundation\\[2ex]

\textbf{The \emph{CoRisk}-Index online:}\\[0.5ex]
{\notsotiny{\url{www.oxford.berlin/CoRisk}}}\\[2ex]
\textbf{Code and Data}\\[0.5ex]
{\notsotiny{\url{github.com/Braesemann/CoRisk}}}\\[2ex]
\textbf{Version history}\\[0.5ex]
V1.2: 30th March 2020\\[0.5ex]
V1.1: 27th March 2020\\[1ex]
All versions available on:\\[0.5ex]
{\notsotiny{\url{arxiv.org/abs/2003.12432}}}}
\runningauthor{The \textit{CoRisk}-Index}
\begin{document}
\begin{frontmatter}
\maketitle
\begin{abstract}
\small While the coronavirus spreads, governments are attempting to reduce contagion rates at the expense of negative economic effects. Market expectations plummeted, foreshadowing the risk of a global economic crisis and mass unemployment. Governments provide huge financial aid programmes to mitigate the economic shocks. To achieve higher effectiveness with such policy measures, it is key to identify the industries that are most in need of support.

In this study, we introduce a data-mining approach to measure industry-specific risks related to COVID-19. We examine company risk reports filed to the U.\,S. Securities and Exchange Commission (SEC). This alternative data set can complement more traditional economic indicators in times of the fast-evolving crisis as it allows for a real-time analysis of risk assessments. Preliminary findings suggest that the companies' awareness towards corona-related business risks is ahead of the overall stock market developments. Our approach allows to distinguish the industries by their risk awareness towards COVID-19. Based on natural language processing, we identify corona-related risk topics and their perceived relevance for different industries. 

The preliminary findings are summarised as an up-to-date online index. The \emph{CoRisk}-Index tracks the industry-specific risk assessments related to the crisis, as it spreads through the economy. The tracking tool is updated weekly. It could provide relevant empirical data to inform models on the economic effects of the crisis. Such complementary empirical information could ultimately help policymakers to effectively target financial support in order to mitigate the economic shocks of the crisis.  

\end{abstract}
\end{frontmatter}

\section{Introduction}
With COVID-19 (``coronavirus'') reaching the level of a pandemic, governments and companies around the world are also exposed to the resulting risks for the highly interconnected global economy. To slow down the spread of COVID-19, governments in China, Europe, the US, and beyond are taking drastic measures, such as travel warnings, border and store closures, regional lockdowns, and curfews.

\begin{figure}[!t]
\centering
\includegraphics[width = 1\linewidth]{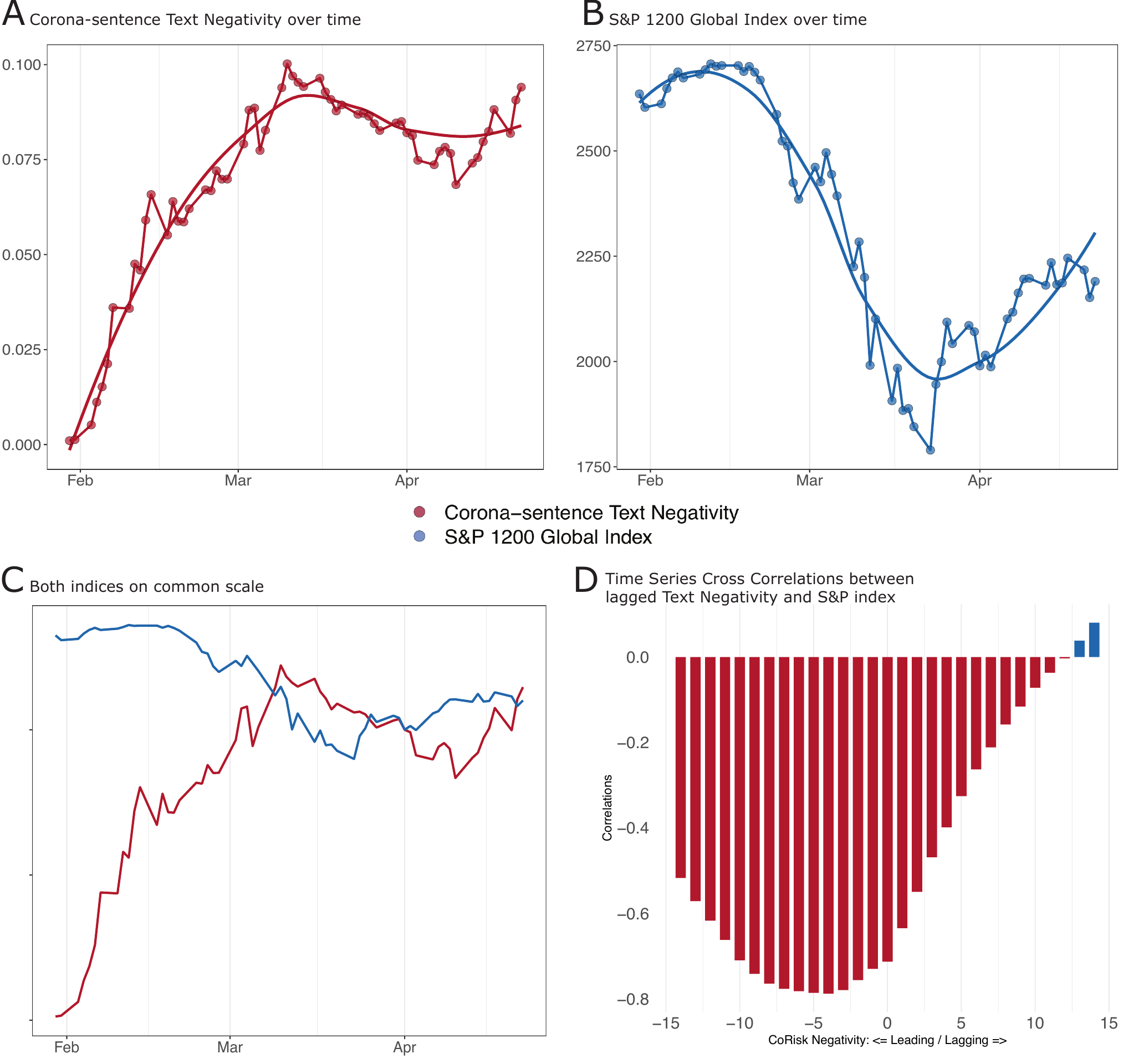}
\captionsetup{width=1\linewidth}
\caption{\footnotesize{\sf{(\textbf{A}) Text 'Negativity' (sentiment) in sentences mentioning coronavirus of 10-k reports. (\textbf{B}) S\,\&\,P Global 1200 Stock Index from 30/01\,--\,23/04/2020. (\textbf{C}) Both time series from panel A on a common scale: they seem to be negatively correlated (\textbf{D}) Time series cross-correlation plot of both time series: the correlations between the corona-sentence text negativity and the S\,\&\,P Global 1200 index are highest (around $\rho = -0.7$) for a S\,\&\,P-lag of around 4 to 7 days; in other words: the CoRisk negativity score leads, indicating a forward-looking risk-awareness of firms with regards to the economic consequences of the pandemic.}}}
\label{fig:fig1}
\end{figure}

These measures are having drastic consequences for personal freedom and the economy. Some sectors, such as airlines or hotels, are facing a nearly complete breakdown of demand. As a first reaction to the expected economic shocks, the global stock markets have collapsed (see Fig.\,\ref{fig:fig1}B). In an attempt to mitigate the general economic downturn, governments all over the world are providing considerable financial support. The US government, for example, is preparing an aid package of 2.2\$ trillion in response to the virus \cite{werner_house_2020}. The German government plans to take up to 156 billion Euros of additional debt (equivalent to half of the federal budget for 2020) to support the economy \cite{zacharakis_wirtschaftsmasnahmen_2020}.

Many of these immediate aid packages are not targeted to specific industries, but are meant to support businesses in all parts of the economy. While such general programmes can help to stabilise financial markets in the short term, it is paramount for their long-term effectiveness to concentrate the support to those areas of the economy that are most in need. For this reason, governments have to identify the industries that are most severely affected by the coronavirus pandemic. However, in the current situation, policymakers lack reliable and up-to-date empirical data, which would allow to assess industry-specific economic risks in real-time. Such information would be crucial to effectively target financial support as the crisis hits the economy and to mitigate the economic shocks.

The study presented here investigates a potential data source that could provide an empirical basis to identify industry-specific economic risks related to COVID-19 and to inform models on the economic effects of the current crisi. We examine company risk reports (10-K reports) filed to the U.S. Securities and Exchange Commission (SEC) and introduce a data-mining approach to measure firms' risk assessments.\footnote{SEC filings represent financial statements of publicly listed companies including a risk assessment. SEC filings are imperative to comply with legal and insurance requirements and therefore should contain the most relevant risks. As a result, companies have a strong incentive to neither under- nor overestimate risks. Moreover, analysing the most recent two-month period of SEC filings represents a random sample of all companies since companies are obliged to report to the SEC at a fixed, but randomly assigned date, independent of their industry.} In collecting all reports published from 30th January 2020\,---\,the day the term \emph{coronavirus} first appeared in a 10-K report\,---\, we can assess and track the reported risk perceptions related to COVID-19 for different industries.

Preliminary findings suggest that the company risk reports' show a forward-looking awareness of potential economic risks associated with the corona-crisis, which is leading stock market developments. Moreover, the awareness towards COVID-19 differs substantially between industries. For example, while 78\,\% of the firms in retail have mentioned the coronavirus as a potential economic risk, only 23\,\% of the businesses in financial services have done so. Thirdly, based on natural language processing, we can identify specific corona-related risk topics and their perceived relevance for the different industries. Lastly, our approach allows us distinguish the industries by their reported risk awareness towards COVID-19. The empirical information provided could help to inform macro-economic models on the effects of the corona-crisis \cite{del_rio-chanona_supply_2020,beland_short-term_2020,ludvigson_covid19_2020} and, thus, help to inform policymakers to better target current economic support programmes to industries that report most severe risks at the current phase of the crisis.The preliminary findings presented here are summarised in one compound index. The \emph{CoRisk}-Index\footnote{\url{http://oxford.berlin/CoRisk}} tracks industry-specific risk assessments related to COVID-19 in real-time. It is available on an interactive online dashboard.

\newpage

This tool is constantly updated. It allows researchers, policymakers, and the public to estimate potential risk factors of COVID-19 in individual sectors of the economy. As the risk filings database is updated on a daily basis, the online tool will allow tracking the potential impact of the crisis as it unfolds and spreads through the economy. With more firms providing risk reports that describe the potential or real impacts of the crisis, the tool will be refined to allow for a more granular perspective on individual industries and sub-sectors.

\newpage

\begin{figure}[!t]
\centering
\includegraphics[width = 1\linewidth]{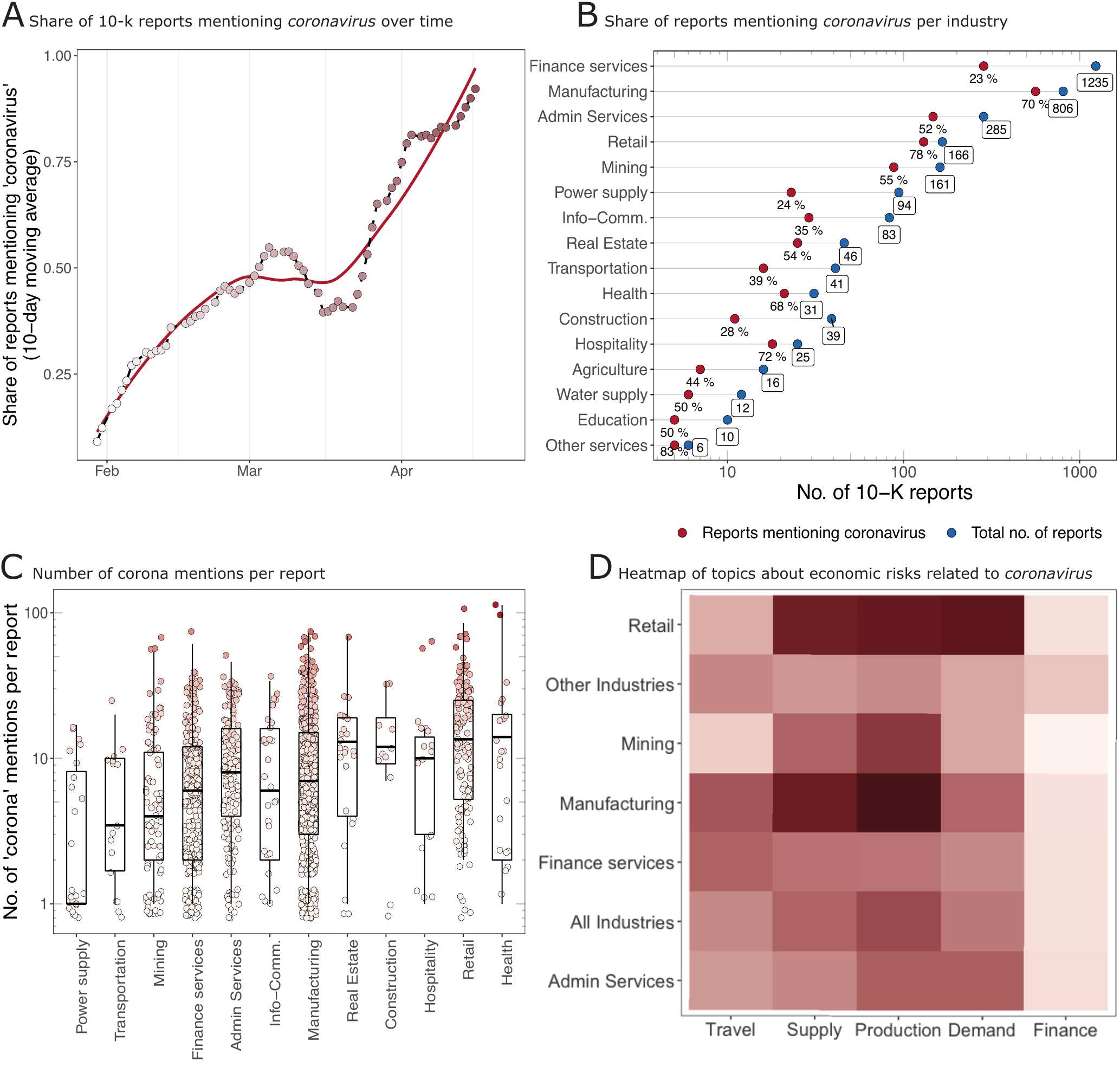}
\captionsetup{width=1\linewidth}
\caption{\footnotesize{\sf{(\textbf{A}) Share of 10-K reports mentioning \emph{coronavirus} over time: with the global spread of the crisis, the share of firms that mention the pandemic as a risk factor increases sharply.(\textbf{B}) Number of 10-K reports filed since 30/01/2020 and share of reports mentioning \emph{coronavirus}. (\textbf{C}) Number of \emph{coronavirus} mentions per report.  (\textbf{D}) Heatmap of relevant economic topics mentioned in 10-K reports related to \emph{coronavirus}.}}}
\label{fig:fig2}
\end{figure}

The study is structured as follows: The next section introduces relevant related research focusing on the economic consequences of infectious diseases and risk assessment in highly interconnected economies. In section \ref{sec:methods} we discuss our data and methods before presenting preliminary results in section \ref{sec:results}. Lastly, we discuss policy implications, methodological limitations, and planned extensions of our approach. 

\begin{figure}[!t]
\centering
\includegraphics[width = 0.8\linewidth]{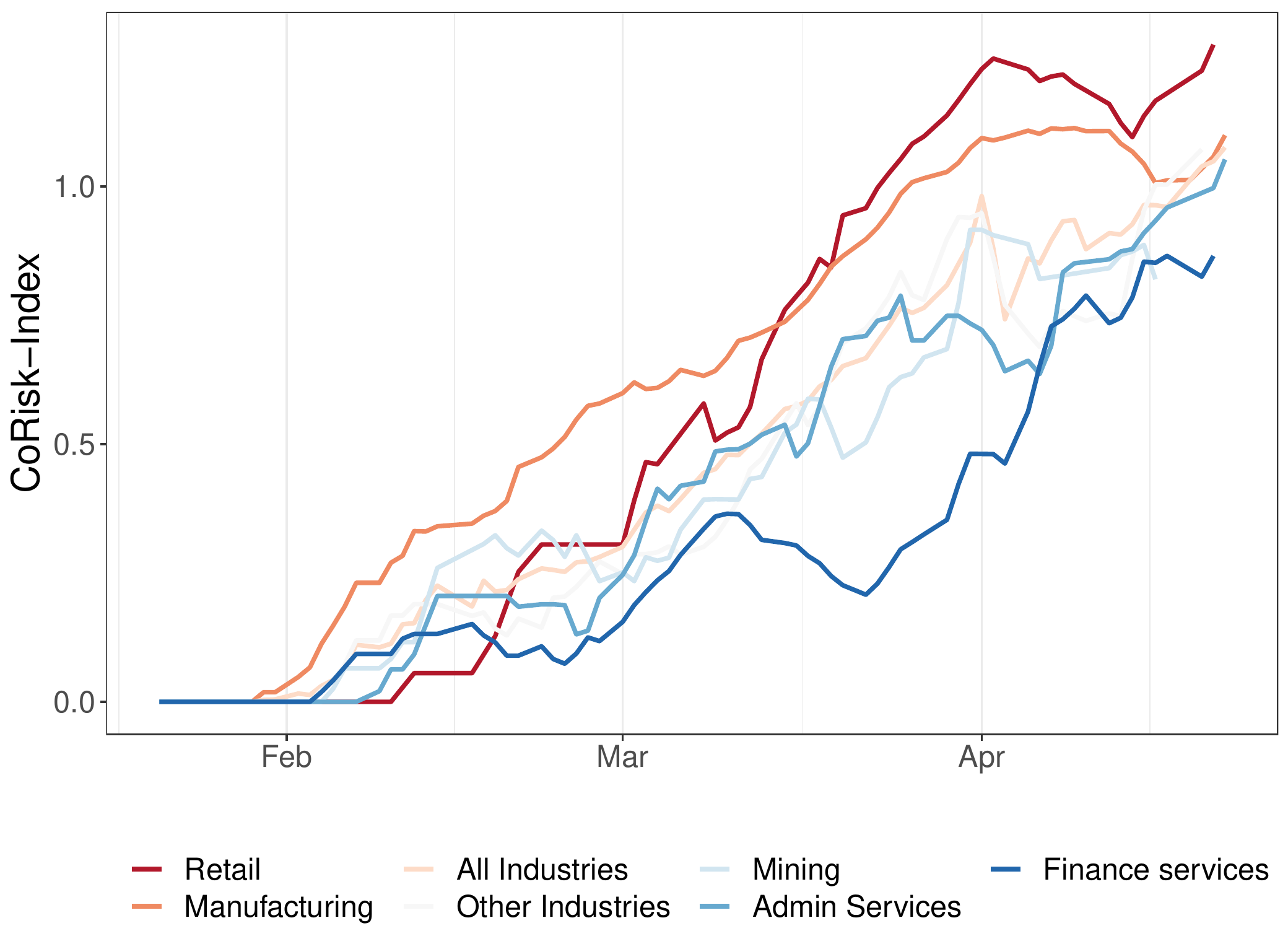}
\captionsetup{width=1\linewidth}
\caption{\footnotesize{\sf{ The \textit{CoRisk}-Index: the index is updated weekly. It is a compound measure of the share of corona-mentioning firms per industry,the average number of corona-mentions per report and the industry-specific corona-sentence text negativity.}}}
\label{fig:fig3}
\end{figure}

\section{Related Work}\label{sec:background}
In this section, we review related work on the economic consequences of COVID-19 and studies on previous epidemics. Moreover, we discuss the assessment of economic risks via reports such as the 10-K reports required by the U.\,S. Securities and Exchange Commission (SEC).

\subsection{Studies on the economic effects of COVID-19}
Global pandemics of infectious diseases are not a new phenomenon. Throughout the last century, the world experienced several global and regional outbreaks: Most severely, millions died during the spread of the Spanish flu in 1918-1920. More recently, smaller epidemics spread around many countries, such as SARS in 2002, the swine flu in 2009 and Ebola in 2014. Nonetheless, the current coronavirus pandemic is, in recent history, unprecedented in its global social and economic consequences. Governments are taking drastic measures while researchers attempt to provide urgently needed policy advice. Much of the coronavirus-related social science and public health research focuses on disease transmission, global spread, and different interventions (see for example \cite{wu_estimating_2020,bogoch_potential_2020,ferguson2020impact,fong2020nonpharmaceutical, nicolaides_hand-hygiene_2019}).

A rapidly growing body of literature investigates the (potential) economic consequences induced by the COVID-19 pandemic \cite{del_rio-chanona_supply_2020,beland_short-term_2020,ludvigson_covid19_2020,lewis_us_2020,baldwin_mitigating_2020,dorn_volkswirtschaftlichen_2020}.\footnote{At this point we want to highlight that it is not the aim of this study to provide a model of risk forecasting that is competing with established macro-economic approaches as described in this section. In contrast, the data-driven methodology presented here aims to explore an alternative data source that could help to inform such economic models. The advantage of the data we provide here is the high time resolution. More traditional sources of empirical information used to calibrate macro-economic models usually include, for example, unemployment rates. While the value of such statistics is undisputed, they are reported with a time-lag. We perceive the purpose of the index provided here to be a complementary data source that could be compared with official statistics on the economic effects of the crisis over time.} Many of these studies aim to assess the potential economic consequences in presenting simulation-based macroeconomic models. For example, \citet{dorn_volkswirtschaftlichen_2020} use scenario calculations to estimate the economic costs of the pandemic for the case of Germany. The authors estimate the financial consequences for the state budget and the employment effects depending on the length of the economic shutdown. Moreover, they include differential adverse effects by sectors, based on press releases and the provisional Ifo business climate index for March 2020. They conclude that the travel and restaurant industry is likely to face a complete shutdown, whereas the pharmaceutical, logistics and health sectors are likely to continue to operate at full capacity. \citet{ludvigson_covid19_2020} aim to estimate the macroeconomic consequences of the pandemic in investigating the impact of disasters in the recent U.\,S. history. \citet{baldwin_economics_2020,baldwin_mitigating_2020} collect a great variety of perspectives on the potential economic implications of COVID-19. Topics range from impacts on trade, economic policy measures, monetary policy, and finance to labour market effects. These contributions rely on simulations, scenarios, descriptive statistics and qualitative arguments.

Regarding the economic impact by sector, \citet{gopinath_limiting_2020} identifies manufacturing and the services sector as disproportionately affected in China, based on the Purchasing Managers’ Index. \citet{ramelli_what_2020} uses Google search intensity to measure attention paid to COVID-19 and stock market data to reveal the economic impact by sector. Their analysis shows that the energy, retail, and transportation sector experienced the largest losses in the United States and China, whereas health care gained considerably in both countries. The analysis by \citet{huang_saving_2020} (also based on stock market data) confirms that the services sectors seem to be the most severely affected in China, and \citet{del_rio-chanona_supply_2020} provide quantitative predictions of first-order supply and demand shocks to the U.\,S. economy on the level of individual industries.

In sum, the contributions presented in this section provide valuable insights, quantitative scenario calculations, and timely policy recommendations. Most of the  macroeconomic analyses are, however, based on assumptions-driven simulations and models, as up-to-date empirical data to assess the immediate economic consequences or industry-specific risks are lacking. These studies could benefit from the alternative data set we are exploring in this study.

\subsection{Historical pandemics}
While research on recent epidemics are limited to simulations \cite{keogh-brown_macroeconomic_2010,buetre_avian_2006}, or specific sectors \cite{rassy_economic_2013}, the study of historical pandemics might provide informative empirical assessments of pandemic-related economic effects. Studies on the 1918 Spanish Flu confirm the primordial effectiveness of non-pharmaceutical (such as, for example, social distancing) interventions even if these come at the cost of economic slowdowns \cite{hatchett_public_2007,morse_pandemic_2007}. Moreover, based on the historical data, researchers find a correlation between mortality rates and declines in GDP, consumption and returns on stocks \cite{barro_coronavirus_2020} and an increase in poverty \cite{karlsson_impact_2014}. However due to the global scope of the crisis for the highly inter-connected world economy of 2020, insights derived from past epidemics are of limited use in identifying the various industry-related risks during the COVID-19 pandemic.

Both, the research on COVID-19 and the historical pandemics rely on stock market information to quantify the economic effects of infectious diseases. However, stock market information comes with several drawbacks. Most importantly, stock markets are prone to irrational herd behaviour and prices capture a variety of information signals into one aggregated index. Examining current stock market dynamics reveals a general economic downturn, but it does not allow to isolate the sector-specific COVID-19 risks. Therefore, we propose to use SEC reports which include risk statements. We argue that these reports represent a promising real-time measure of industry-specific business risks. Furthermore, the analysis of report statements discussing 'coronavirus' allows to isolate the business risks exclusively associated with the COVID-19 outbreak.

\subsection{Assessing economic risks via business reports}

Since the great recession hit the world economy in 2008, risk has been a crucial topic in governance and finance. While risk assessments of the financial system led to diverse measures to make the world economy less vulnerable to economic shocks originating in the financial sector, a health crisis, such as the current pandemic, poses different risks to the economy. While government measures against the spread of the disease hinder the population from working and consuming, which results in businesses interrupting production, many economies face demand and supply shocks at the same time. In particular, as different industries rely on distinct input factor compositions and supply chains, the sectors of the economy react differently to shocks \cite{bogataj_measuring_2007}. Regarding the COVID-19 crisis, we expect sectors whose operations are more connected to supply from manufacturers in China to publicly report corona-related risk earlier than others. These sectors are also highly connected and interdependent within the national economy and risk might spread between sectors \cite{atalay_network_2011, acemoglu_network_2012}. 


Most risk assessment approaches focus on quantitative probability-based methods and financial data \cite{aven_risk_2012, aven_foundational_2019}. The data published in such quantified risk assessments is often made available retrospectively, which makes a real-time evaluation of risks difficult. In contrast to such assessments, we investigate the annual 10-K reports filed to the U.\,S. Securities and Exchange Commission (SEC), which provide verbal corporate risk disclosure and financial statements. Besides, SEC filings are imperative for legal and insurance requirements and they need to contain the most relevant risks to protect the company from legal liabilities. Depending on the volume of publicly traded stocks, companies with a public float of over 700 million USD are obliged to report within 60, while smaller companies have 75 or 90 days after the end of the fiscal year. Furthermore, the reports inform investment decisions and risk governance at the same time and, thus, companies, as rational agents, are likely to communicate moderate risk assessments \cite{richman_sec_2019}. In fact, prior work has underlined the forward-looking nature of the reports, since they allowed a more effective prediction of volatility on stock market returns than the compared approaches \cite{kogan_predicting_2009}. Correspondingly, we expect the 10-K reports to also provide forward-looking information on risk assessments during the observed time period and in particular on rising business risk in relation to the spread of COVID-19.



Moreover, the filing companies are categorised into sectors using the "Standard Industry Classifier", which allows a real-time analysis of industry-specific risk developments. In this study we apply natural language processing to extract corona-related risk information from the reports and analyse sector-specific differences in risk awareness and disclosure (details in section \ref{sec:methods} and in the appendix \ref{sec:appendix}).

\subsection{Hypotheses}
 
Based on the literature on the potential economic effects of pandemics and on the information available in business risks reports, we derive the working hypothesis\footnote{While the results provided in this working paper are largely descriptive, we present the research hypotheses, which have guided our data exploration.} of this study:

\begin{quote}
\textit{\underline{Working hypothesis:} SEC 10-K reports contain corona-related information, which allow to track the industry-specific economic risk assessments in near real-time as the economic crisis unfolds.}
\end{quote}

This working hypothesis is split into three operationalised hypotheses. A core assumption of the investigation of the risk reports is that they contain economic meaningful information. Accordingly, the sentiment of the reports should reflect overall market conditions:

\begin{quote}
$H1$\textit{: 10-K risk reports contain economic relevant information on short-term market expectations, i.\,e. they are correlated with overall stock market trends in the current crisis.}
\end{quote}

In contrast to stock indices, which are essentially a highly aggregated signal of all market expectations, risk reports provide a detailed outlook of individual risk factors and their potential influence on business outcomes.\footnote{This should not imply that the firms have a correct risk forecasting model in place or that they actually use all available information to provide a best-possible risk prediction. Instead, we assume that the firms report risks that they perceive as relevant at the time of reporting (see also the methodological limitations section below). A presumption of our approach is that firms try to 'honestly' report relevant risks, not that they actually predict them. With more data becoming available over the next weeks (in particular first statistics on the macro-economic effects, such as unemployment rates), we will incorporate predictive models to get a better understanding in how far the reported risk factors indeed correlate with adverse effects experienced.} As some of the potential economic effects of the corona-crisis could be already foreseen when the disease was still largely concentrated in China (in particular risks with respect to supply chain interruptions), we hypothesise that:

\begin{quote}
$H2$\textit{: Corona-related risk assessments reported in 10-K filings show a time lead compared to stock market dynamics.}
\end{quote}

Industries are affected differently during the crisis, depending on their business model. As introduced in \cite{benassy-quere_covid-19_2020}, the economic crisis will unfold in several phases with different characteristics. Those sectors of the economy that are more vulnerable to supply chain interruptions and short-term collapses of consumer demand should be more affected in the early phase of the crisis than other sectors:

\begin{quote}
$H3$\textit{: Corona-related risk factors in 10-K reports differ between industries.}
\end{quote}





To test the hypotheses, we use web-mining techniques to collect data from SEC 10-K reports and conduct text mining to extract information relevant for the individual hypotheses, as outlined in the next section. Section \ref{sec:results} presents the results of the analysis.

\section{Methods}\label{sec:methods}
The 10-K SEC filings, as legal reports to publicly communicate corporate risks and financial statements, provide a valuable and innovative text data source for risk assessment.\footnote{The data source explored here is not meant to replace any established macro-economic statistics, but rather to provide a complementary alternative source of data to identify immediate industry-specific risk perceptions.} 

We use a web scraper to collect all 10-K filings published since 30/01/2020 from the SEC's "Electronic Data Gathering, Analysis and Retrieval System" (EDGAR) database. The filing documents contain the company names and central index keys (CIK) as unique firm identifiers and the Standard Industry Classifier (SIC) that allow linking and comparing the filings data with individual business and industry data. Each 10-K filing contains a Risk Factor section (Item 1A) under which the reporting company is obliged to disclose all types of risks their business might be facing to adequately warn investors. Companies are required to use "plain English" in describing these risk factors, avoiding overly technical jargon that would be difficult for a layperson to follow. The text of each 10-k report builds the document by which a company is represented in our analysis. The entire text is set to lower case for further analysis, before the occurrence of the main two keyword tokens, \textit{'corona'} and \textit{'covid'}, is counted. Any word containing one of the two main tokens is likewise counted. Hereby the measure 'corona-keyword count per report' is created. Similarly, each company that reports one of the keywords at least once is included in the 'share of corona-mentioning firms per industry'. 

After text pre-processing, we apply different Natural Language Processing tools to analyse the reports. Different sectors are facing different challenges, therefore companies are reporting about different corona-related risks. We aim to capture these risk topics via a keyword search on predefined topics. In order to explore possible topics, we used Latent Dirichlet Allocation (LDA) for unsupervised topic modelling, similar to \cite{dyer_evolution_2017}. We only apply the topic model to corona-related paragraphs in the risk sections. We additionally examine the most frequent words and bi-grams in the documents. Using this exploratory analysis, we define a set of topics, which are specified by keywords. We then conduct a keyword search to count how much these terms are mentioned in the different industries in order to estimate the topic prevalence. The resulting topic heatmap (see section\,\ref{sec:results}) reports the share of sentences per topic per 1,000 corona-containing sentences for the different industries. Moreover, we measure the sentiment of corona-related sentences via the share of negative words \cite{loughran_when_2011}. A more detailed description of the different methods can be found in the appendix \ref{sec:appendix}.


\section{Preliminary Results}\label{sec:results}

The main results of our analysis are displayed in Figure \ref{fig:fig1} as well as in Figure \ref{fig:fig2}.\footnote{At this point we want to highlight, again, that the results reported here are preliminary. With more data becoming available over the next weeks and, hence, with a refined methodology, we expect to be able to report more in-depth analyses and conclusive findings. These will be constantly updated on the online dashboard.\\
Moreover, we want to emphasise that all quantitative findings are build on data of U.\,S. firms. It remains to be seen in how far the findings could be extrapolated to other economies.} The overall sentiment of the 10-K reports becomes more negatively sharply (Fig.\,1A), closely related to the collapsed stock markets after 15th February (Fig.\,1B). This timely correlation provides evidence in favour of research hypothesis $H1$\,---\,the 10-K reports provide economically relevant information.

Moreover, the sentiment of the text sections that discuss the corona-related business risk factors decreases already a few weeks before the overall negative business outlooks manifested in falling stock prices (Fig.\ref{fig:fig1}C). This observation supports hypothesis $H2$: Corona-related risk sentiments show a time lead compared with overall stock market dynamics. Indeed, this visual evidence is corroborated by the cross-correlations displayed in Fig.\,\ref{fig:fig1}D. The figures compares the corona-sentence text negativity and lags of the S\&P Global 1200 index. The correlations are highest with a lag of four to seven days, indicating that the text negativity score leads. From this comparison, we conclude that the 10-K reports contain business relevant information that can be isolated\,---\,in contrast to the aggregated signal covered in stock indices alone. Investigating the reports can, thus, provide additional information about the firm- and sector-specific risks associated with the pandemic.

Differences in corona-related risk assessment by industries are displayed in Figure \ref{fig:fig2}. Overall, From the end of January 2020, an increasing number of 10-K reports refers to the coronavirus (Fig.\,2A), indicating an increasing awareness of COVID-19 as a potential economic risk. Not all sectors of the economy show a similar awareness of the potential business impacts of the pandemic (Fig.\,\ref{fig:fig2}B), which provides evidence in favour of hypothesis $H3$. For example, while 70\,\% of the firms in manufacturing mention corona-related risks, only 23\,\% of the firms in the financial sector consider such risk factors. Other sectors that show a high awareness of corona-related risks are retail (78\,\%), or hospitality (72\,\%).\footnote{In some of the industries, only very few firms have provided risk reports in the current observation period. In particular in these groups, the results are likely to change as more data become available.}.

The firms, moreover, differ substantially with regards to the intensity with which they discuss the potential impacts of the coronavirus to their businesses in the 10-K reports. As Fig.\,\ref{fig:fig2}C shows, some firms, in particular in retail and manufacturing, mention terms related to the pandemic (\emph{'corona'}, \emph{'coronavirus'}, \emph{'COVID'}) substantially more often than other firms. 

The 10-K reports moreover provide information about specific types of risk perceptions. Fig.\,\ref{fig:fig2}D provides a heatmap with important topics per sector. The rows represent five relevant topics (demand, finance, production, supply, and travel), which have been derived from unsupervised topic modelling and subsequent uni- and bi-gram search.\footnote{Details of the text mining approach and the final keywords that define the topics are provided in the appendix.} The cells of the heatmap are coloured according to the topic relevance, i.\,e. the number of topic-specific keywords per industry. Additionally, we have applied a hierarchical clustering algorithm on the data to identify related topics and sectors, which report about similar topics ( not displayed in the figure).


The figure reveals that most firms either report demand and financial risk factors 
or production and supply chain risks.
For example, supply chain problems represent the biggest reported risk component for firms in the retail industry, while manufacturing firms report both supply and production risks, as well as finance and potential demand risks. While reports of the mining industry consider demand-related issues as the biggest risk factor, other sectors
do not report extensively about any of the specific risk types. 

These findings support hypothesis $H3$\,---\,corona-related risk factors reported in 10-K reports differ between industries with regard to occurrence and topical context. The reports are, thus, a valid data source to identify sectors that face particular risk factors in the current early phase of the crisis.

While a one-dimensional categorisation of risk assessments tends to over-simplify the crisis firms are facing, it allows to compare the different industries and to identify those parts of the economy, which currently report more or less severe effects due to the immediate economic consequences of the pandemic. The data-driven assessment reveals, in particular, that manufacturing and retail are among the industries that report to be most vulnerable to the changing economic environment. Not only decreasing consumer demand but particularly problems along the supply chain mark substantial risk factors for those industries. On the other hand, it is mostly firms from the information and communication services sector, which are less dependent on the physical transport of goods, that are currently reporting fewer risks. Moreover, with only 28\,\% of the 10-K reports mentioning the coronavirus as a potential risk factor, the construction sector seems largely unaffected by the short term economic shocks.

The extent to which the industries are affected by corona-related business risks is likely to change over time. Accordingly, the risk categorisation presented here is a static measure that needs constant updating and refinement as the crisis unfolds. To do this, the study is supplemented by an online dashboard, which tracks the main findings with respect to the risk categorisation over time. A static visualisation of the tracking tool is displayed in Fig.\,\ref{fig:fig3}. Due to limited data availability, the time-tracking could, so far, only be conducted for the largest industries; all other industries are aggregated in 'other industries'. The \emph{CoRisk}-Index is a compound measure (i.\,e. geometric mean) of the share of firms that have reported corona-related risks (see Fig.\,\ref{fig:fig2}B) the average number of corona-keywords per report (see Fig.\,\ref{fig:fig2}C) and the industry-specific text negativity (see Fig.\,\ref{fig:fig1}A), aggregated weekly. The different industries can be dinstiguished by their perceived risks they are reporting.
With more reports being released over time, we will update and refine the \textit{CoRisk}-Index to allow for more fine-grained industry-specific analyses. The index starts with zero, indicating no corona-related business risks in the reports. We will track the \textit{CoRisk}-Index over the course of the corona crisis until the index reaches zero again.


\section{Conclusion}

\subsection{Summary}

As the COVID-19 pandemic unfolds, governments are attempting to reduce contagion rates at the expense of personal freedom and negative economic effects. Sizeable cyclical and fiscal policy packages are prepared in order to counterbalance the dooming global economic downturn. In order to ensure an effective usage of public crisis spending, it is paramount to understand in detail which industries are most affected by the pandemic and currently most in need of support. This study introduces a data-mining approach to measure the reported business risks induced by the current COVID-19 pandemic. We examine company risk reports filed to the U.S. Securities and Exchange Commission (SEC). Harnessing this data set enables a real-time analysis of potential risk factors. Preliminary findings show that the companies' risk awareness is preempting stock market developments. While stock prices typically condense the market's multiple signals, the 10-K reports allow to isolates the company risk perceptions associated to the COVID-19 outbreak. Moreover, this risk awareness differs substantially between industries, both in magnitude as well as in nature. Based on natural language processing techniques, we can identify specific corona-related risk-topics and their relevance for the different industries: supply chain- and production related issues seem to be mostly relevant for retail and manufacturing, while several industries have reported demand- or finance related risk factors. 
We summarise the corona-related business risk perceptions per industry in a compound index published online.\footnote{\url{http://oxford.berlin/CoRisk}} The \emph{CoRisk}-Index, which will be updated constantly over the course of the crisis, provides an up-to-date database to identify the industries that report most substantial risk factors in the different phases of the unfolding economic crisis. The online dashboard can provide a data source to inform economic models and to provide empirical information that could help to assess potential policies, which aim to effectively target financial support and to mitigate the economic shocks of the current crisis.  

Governments are eager to counterbalance the dooming global economic crisis induced by the COVID-19 pandemic with cyclical and fiscal policy packages of enormous volumes. Democratic accountability demands this public crisis stimulus to be spend as effectively as possible. Our data-driven analysis of the 10-K SEC filings provides an alternative data source, which could be used to calibrate macroeconomic models and thus help identifying industries that are likely to be most severely affected by COVID-19. In particular the manufacturing and retail industries are currently reporting most severe effects from the sharply decreasing consumer demand and interruptions of the supply chain, and might face substantial problems through the unfolding of the crisis.\footnote{At this point, we want to emphasise again that the findings are preliminary and might change over time, as more data becomes available and as the research methodology is refined. Moreover, the risk categorisation presented in this study is not meant as a direct policy recommendation. Instead, the tool is meant to provide an empirical source of information for real-time tracking of industry-specific risk perceptions, which can help to inform policy makers in times of a fast-evolving economic crisis.} 

Other parts of the economy, in particular information-processing service industries, are currently reporting less severe corona-related risks. However, as the shock will transmit throughout the tightly inter-connected economy over time, these industries are likely to face larger challenges at a later stage in time. Nonetheless, at the current stage, their core businesses seem to be less directly affected by supply chain interruptions and collapsing consumer demand than other industries. 

\subsection{Methodological Limitations} 
Our approach is based on the risk assessments in the U.\,S. Securities and Exchange Commission (SEC) filing reports. Thus, the value of the approach relies crucially on company self-reporting. While the firms are unlikely to provide a risk prediction with a high forecasting accuracy, it might still be worth exploring the reports as alternative data source to measure risk perceptions. As the reports serve as legal and insurance requirements against financial risks, but also as a basis for investment decisions of investors, companies are implicitly encouraged to neither over- nor understate the risks they are facing. Nevertheless, our results are limited to this self-assessment. As many of the implications of the corona crisis are still uncertain, our approach thus reflects a way to approximate potential implications on current estimations of experts in the different sectors, represented by the companies, and does not include risks that are unforeseeable for themselves at a given point in time. Moreover, the data to assess the precision of these estimations does not exist yet, as there has not been a pandemic with comparable global economic consequences in recent time. In the future, we will evaluate our results by looking at employment data in different sectors. Moreover, the pandemic continues to spread and will soon affect all industry sectors and all countries of the world. As more and more companies will report on related risks, the count of corona mentions alone will lose its information-value as a measure to differentiate between endangered industry sectors. Then, more granular measures, such as the identified topic categories, will become more important to distinguish between different natures of risks.

The exploration of alternative data sources that are meant to complement or now-cast established economic statistics always comes with uncertainty. For example, it is not yet clear how the short-term risks described by the different industries will translate into long-term economic outcomes, such as bankruptcies. Nonetheless, we believe that the reports could be a reliable source of empirical information about the issues faced by different industries in the current situation, and they might be used to inform forecasting models on industry-specific economic effects of the crisis, as they help fill a data gap. Models that incorporate alternative data sources such as the one presented here could then be beneficial for developing economic support packages that are currently provided by governments. 

All technical methods serve the higher purpose of providing timely and comprehensible insights into the industry-specific effects caused by the global outbreak of the coronavirus. To mitigate susceptibility to errors and increase reproducibility, we mostly draw from more basic technical methods. This can be seen in the discovery of Corona-relevant keywords which is based on the matching of regular expressions to avoid error-prone text pre-processing. Reduction of technical complexity, however, comes at the cost of diminished modelling fidelity and potential accuracy of results. For instance, the sentiment classifier has not been fine-tuned on text snippets discussing financial risks in particular. Further, it may be questioned whether a generalist sentiment score yields a reliable measure for the assessment of risks. Additionally, the LDA-based topic modelling approach lacks in interpretability and robustness, and we therefore only use it for exploration until now. The topics and keywords we use for estimating topic prevalence are therefore hand-coded which limits the detection of topics to predefined terms.

In general, the findings presented here should be considered as preliminary. Sensitivity checks are needed to validate their robustness, and numerous extensions will help to better assess the potential value of the exploration of 10-K reports as a complementary data source to measure industry-specific risk factors.

We expect some of these limitations to be mitigated with more reports to be analysed in the near future. Adjusted results will be published on the online dashboard and in a refined version of the working paper. 

\subsection{Future Work}
The following impetus for future work builds directly on the identified methodological limitations and caveats mentioned throughout the paper. Rather than harnessing generalist sentiment analysis, future efforts should explore risk classifiers trained on labelled financial risk assessments and entity sentiment analysis \cite{loughran_when_2011,falck_measuring_2019,stoehr_mining_2020,darius_hashjacking_2019}.

The robustness of the results needs to be checked in more detail. In particular, we will compare historical and unemployment data with risk measures (text negativity) extracted from 10-K reports to investigate the correlation between risk reports and overall economic trends. Moreover, we will weight the index by firm size and compare the composition of the index with the overall U.\,S. economy in order to investigate the representativeness of the sample.


Likewise, the content of other online platforms, such as Wikipedia, as a source for risk assessment could be explored \cite{stephany_exploration_2017,stephany_coding_2019}. Disentangling international cooperation patterns \cite{braesemann_global_2019,stoehr_disentangling_2019} and mining industry-specific key technologies \cite{stoehr_mining_2020} will be highly beneficial. This could allow an estimation of operational risk channels and risk spillovers propagating between industries and countries due to global supply chains and peer-tier co-operations. Considering more financial data could include credit risks (financial exposures) of companies. Besides the comparison with additional financial signals, we plan to include unemployment rates in our analysis as an exogenous variable to the SEC risk assessment. 

\section*{Acknowledgements}
We wish to thank Andrew Stephen, who organised a rapid informal peer review process at the Sa\"id Business School Oxford, and Felix Reed-Tsochas and Felipe Thomaz who served as faculty internal referees. Their extensive and valuable feedback on V1.1 helped to substantially improve the working paper and to more clearly point out the contribution of the study.

Moreover, we would like to express our gratitude to Vili Lehdonvirta and Scott Hale from the Oxford Internet Institute for their valuable comments. Additionally, we wish to thank Tobias Reisch from the Complexity Science Hub Vienna, Tomaso Duso and the Firms and Markets Department from the German Institute of Economic Research (DIW Berlin), and Slava Jankin and the Data Science Lab from the Hertie School of Governance who gave us the opportunity to present our work. Their feedback on V1.2 has helped us to identify elements of the study that require further work and improvement.

\printendnotes
\renewcommand{\bibsection}{\section*{References}}
\bibliography{cite.bib}

\appendix

\section{Methodological and Technical Appendix}
\label{sec:appendix}

\subsection{Scraping of 10-K SEC Filings}
We use web crawlers and scrapers written in the general programming language Python for finding and extracting the relevant risk section in the 10k reports. Via the SEC online interface\footnote{\url{https://searchwww.sec.gov/EDGARFSClient/jsp/EDGAR_MainAccess.jsp}} past SEC filings are queried since the first emergence of 'coronavirus' in a report on January 30th 2020. A web crawler allows to find the relevant risk factor section in the 10-K reports and stores each paragraph for later text processing and analysis.

\subsection{Collection of Stock Market Data}
The collection of stock market data follows the incentive to provide reference data for the company SEC filings. In both, SEC filings and financial markets, each company is assigned a unique Central Index Key (CIK) and a ticker number. Based on the CIK, we identified a list of 13,737 companies using the data registry provided by Rank and Filed \cite{noauthor_rank_nodate}. This list was then filtered by the list of companies of which we obtained SEC filings. Using the Yahoo Finance API, we finally retrieved stock prices for all remaining companies in the list \cite{aroussi_yfinance_nodate}.

\subsection{Discovery of Relevant Keywords}
\label{sec: Discovery of Relevant Keywords}
Since this study examines the attention attributed to COVID-19 in the SEC filings, the discovery mechanism of relevant COVID-19 mentions is of central importance. To mitigate susceptibility to errors due to word splitting, stemming and other text preprocessing, we decided for the most simple approach based on the matching of regular expressions. We scanned the reports for the two relatively unambiguous terms "corona" and "covid", also accounting for "coronavirus" and "covid-19" without duplication. 

\subsection{Sentiment Analysis on 10-K SEC Filings}
Sentiment analysis on the SEC filings serves the purpose of a more contextualised semantic understanding of the risk assessment concerning COVID-19. We selected the paragraph of the risk report, with the highest number of coronavirus mentions and calculated the sentiment based on the TextBlob API \cite{noauthor_textblob_nodate} using the code developed in \cite{falck_measuring_2019},

\subsection{Topic Modelling on 10-K SEC Filings}
We use unsupervised learning techniques to identify the main topics that companies mention when describing coronavirus-related risks. Latent Dirichlet Allocation (LDA) is a Bayesian computational linguistic technique that identifies the latent topics in a corpus of documents \cite{blei_probabilistic_2012}. This statistical model falls into the category of generative probabilistic modelling: a generative process which defines a joint probability distribution over the observed random variable, i.e. the words of the documents, and the hidden random variables, i.e. the topic structure. In other words, LDA uses the probability of words that co-occur within documents to identify sets of topics and their associated words \cite{dyer_evolution_2017}. The number of topics has to be defined in advance. LDA is a frequently used technique to identify main topics in a corpus. Nevertheless, the interpretation of these topics can sometimes be difficult. We thus perform LDA for explorative purposes in our research only. By additionally exploring the most common words and bi-grams we then define topics and the defining keywords by hand. We then detect how often companies mention these keywords in the corona related risk sections to estimate how important the topics are in different sectors. In particular, we perform the following steps:
\paragraph{Sample restriction}
Referring to section \ref{sec: Discovery of Relevant Keywords}, we filter all sentences from the risk sections that mention either "corona" and "covid", thereby also accounting for "coronavirus" and "covid-19".
\paragraph{Text preparation}
Before we train the LDA model we prepare the documents to achieve better performance of the method. We remove all common English stopwords, which are frequent words such as “is,” “the,” and “and” as well as those words which appear in at least $80\%$ of the documents. These words are not useful in classifying topics as they are too frequent and therefore decrease performance. Moreover, we delete all words that do not occur in at least $2$ documents. 
\paragraph{Topic modelling with LDA}
We turn the documents into numerical “Bag of words” feature vectors, disregarding word order. We then use LDA to extract the topic structure. Like any unsupervised topic model, this requires setting the number of topics a priori. We selected this key parameter based on semantic coherence, evaluating a range of 2 to 8 topics leading to a final model of 4 topics. The top 10 terms of each topic are displayed in Table \ref{tab:topics}. 
\paragraph{Topic keyword frequencies}
The derived topics give a good insight into the general narratives of risks used in the documents. Nevertheless, they are hard to interpret, as early corona-related risk reports are still generic in that various risk factors are covered. We make use of our insights from the topic modelling to define 5 main topics defined by keywords, displayed in Table \ref{tab:keywords}. The choice of keywords is additionally informed by clustering the most frequently used words and bi-grams in the documents. We then measure the frequency of these keywords per topic per industry in all of the documents to get more insights into the nature of risks different sectors are facing due to the pandemic.

\begin{table}[h]
    \centering
        \caption{Topics and keywords }
    \label{tab:keywords}
    \begin{tabularx}{\linewidth}{c|X}
      \textbf{Topic}   &  \multicolumn{1}{c}{\textbf{Keywords}}\\
      \hline
    Production & business operation, business disruption, product, work stoppage, labor disruption, labor, work, manufacturing operation, labor shortage, employee productivity, product development, business activity\\
Supply & manufacturing facility, manufacture facility, contract manufacturer, service provider, logistic provider, supply disruption, party manufacturer, supply disruption, facility, supply, transportation delay, delivery delay, supplier, business partner, supply chain, material shortage\\
Travel & air travel, travel, travel restriction, airline industry, travel disruption\\
Demand & store closure, distribution channel, market condition, consumer spend, market acceptance, consumer confidence, consumer demand, consume, store, customer, store traffic\\
Finance & operating result, cash flow, stock price, estate value, credit availability, performance problem\\
    \end{tabularx}

\end{table}{}

\begin{table}[h]
    \centering
 \caption{LDA with 4 topics.}
\begin{tabular}{c|c}
 \textbf{Topic number}  & \textbf{Top 10 words} \\
       \hline
Topic 0:&
impact extent including outbreak uncertain future highly results developments depend\\

Topic 1:&
operations including health outbreak business supply products economic public result\\

Topic 2:&
outbreak spread countries impact including china business potential economic government\\

Topic 3:&
china outbreak novel covid adversely wuhan strain business december recent \\
    \end{tabular}
    \label{tab:topics}
\end{table}{}

\end{document}